\documentclass[prl,twocolumn,superscriptaddress,showpacs,floatfix,amsfonts]%
{revtex4}
\usepackage{graphicx,graphics,color,epsfig,exscale}
\usepackage{amsmath,amssymb,bm}

\newcommand{\sgn}{\text{sgn}}
\newcommand{\pdag}{\phantom{\dag}}

\begin{document}
\preprint{}
\title{Finite-Size Scaling of Classical Long-Ranged Ising Chains and
the Criticality of Dissipative Quantum Impurity Models}
\author{Stefan Kirchner}
\affiliation{Department of Physics \& Astronomy, Rice University,
Houston, TX 77005, USA}
\author{Qimiao Si}
\affiliation{Department of Physics \& Astronomy, Rice University,
Houston, TX 77005, USA}
\author{Kevin Ingersent}
\affiliation{Department of Physics, University of Florida,
Gainesville, FL 32611--8440, USA}


\begin{abstract}
Motivated in part by quantum criticality in dissipative Kondo systems,
we revisit the finite-size scaling of a classical Ising chain with
$1/r^{2-\epsilon}$ interactions. For $\frac{1}{2}<\epsilon<1$, the scaling
of the dynamical spin susceptibility is sensitive to the degree
of ``winding'' of the interaction under periodic boundary conditions.
Infinite winding yields the expected mean-field behavior, whereas without
any winding the scaling is of an interacting $\omega/T$ form. The contrast
with the behavior of the Bose-Fermi Kondo model suggests a breakdown
of a mapping from the quantum model to a classical one due to the smearing
of the Kondo spin flips by the continuum limit taken in this mapping.
\end{abstract}
\maketitle

Quantum criticality describes the collective fluctuations of a continuous
quantum phase transition. The standard theoretical picture \cite{Hertz.76} is
that quantum critical fluctuations in $d$ spatial dimensions can be described
in terms of the fluctuations of an order parameter (a classical variable)
in $d+z$ dimensions, where $z$ is the dynamic exponent. However, there
has been much recent interest in quantum critical points (QCPs) that do not
conform to this quantum-to-classical mapping \cite{unconventional_QCPs}.

One setting to explore this issue is quantum impurity models that involve one
or more dissipative bosonic baths having a sub-ohmic ($\epsilon>0$) power-law
spectrum
\begin{equation}
\label{EQ:sub-Ohmic}
\sum_p [\delta(\omega-w_p)- \delta(\omega+w_p)] \sim
   |\omega|^{1-\epsilon} \sgn(\omega) \Theta(\omega_c-|\omega|).
\end{equation}
Here $w_p$ is the dispersion of the bosonic bath(s) and $\Theta$ is the
Heaviside function specifying the high-energy cut-off $\omega_c$. The
quantum-to-classical mapping leads to
a classical Ising chain with $1/r^{2-\epsilon}$ interactions, which has as its
continuum limit a local $\phi^4$ theory in $0+1$ dimension \cite{Fisher.72}.
For $1/2 < \epsilon < 1$, the $\phi^4$ theory has a Gaussian fixed point with
a violation of $\omega/T$ scaling. However, the QCP in a large-$N$,
fixed-$\kappa$ limit of a spin-rotation-invariant SU$(N)\times$SU$(\kappa N)$
Bose-Fermi Kondo model (BFKM) satisfies $\omega/T$ scaling and, hence, is
interacting \cite{Zhu.04}. Numerical renormalization-group (NRG) studies have
also found interacting QCPs for $1/2 < \epsilon < 1$ in two quantum models
with Ising anisotropy: the dissipative spin-boson model \cite{Vojta.05} and
the Ising BFKM \cite{Glossop.05}. The failure of the quantum-to-classical
mapping has been attributed in the spin-rotation-invariant case to the Berry
phase \cite{Kirchner.08unpl}, but remains to be clarified for Ising symmetry.

This Letter reports a two-pronged study of the classical mapping of the
quantum-critical Ising BFKM. First, we show that, for $1/2<\epsilon<1$,
the scaling of the dynamical
spin susceptibility of the classical model as a function of system size and
wave vector (corresponding in the quantum problem to temperature and Matsubara
frequency, respectively) depends crucially on the ``winding'' of the interaction
to extend the range of the interaction beyond the finite system size. For
infinite winding, one recovers the mean-field $\omega/T^{1/(2-2\epsilon)}$
scaling expected from the corresponding $\phi^4$ theory, in which the quartic
coupling acts as a dangerously irrelevant variable. Limiting the winding
introduces an additional temperature dependence dominating that from the
dangerously irrelevant coupling and giving rise for zero winding to $\omega/T$
scaling. Second, we provide strengthened evidence for non-mean-field behavior
in the BFKM. These findings can be reconciled through a breakdown of the
quantum-to-classical mapping arising from failure of the continuum limit to
fully preserve the effects of Kondo spin flips.

\textit{Bose-Fermi Kondo model and the classical Ising chain.}---The BFKM
describes a quantum spin coupled to both a fermionic band and a bosonic bath.
The Hamiltonian for the Ising-anisotropic case is
\begin{multline}
\label{EQ:H-imp}
\mathcal{H}_{\text{BFKM}} = J_K \mathbf{S} \cdot \mathbf{s}_c
  + \sum_{p\sigma} E_p \: c_{p\sigma}^{\dag} \, c_{p\sigma}^{\pdag} \\
+ g \, S^z \sum_p \Bigl( \phi_{p}^{\pdag} + \phi_{-p}^{\;\dag} \Bigr)
  + \sum_p w_p \: \phi_p^{\dag} \, \phi_p^{\pdag} ,
\end{multline}
where $\mathbf{S}$ is a spin-$1/2$ local moment, the $c_{p\,\sigma}$'s
represent a fermionic band with a constant density of states
$\sum_{p} \delta (\omega - E_p) = N_0$ and on-site spin $\mathbf{s}_c$,
and the $\phi_p$'s represent a bosonic bath having the spectrum
in Eq.~(\ref{EQ:sub-Ohmic}). For fixed $N_0 J_K$, a QCP at $g = g_c$
separates Kondo (or delocalized) and local-moment (or localized) phases.
For $g=0$, the BFKM reduces to the conventional Kondo model.

The partition function associated with $\mathcal{H}_{\text{BFKM}}$ is
\cite{Grempel.03+Kirchner.08}
$Z_{\text{BFKM}}\sim \text{Tr} \exp(-S_{\text{imp}}')$, where
\begin{multline}
\label{eff-action}
\mathcal{S}_{\text{imp}}' = \int_0^{\beta} \!\! d \tau
  \biggl\{ \Gamma S^x(\tau) - \frac{1}{2} \int_0^{\beta} \!\! d \tau' \,
  S^z(\tau) \, S^z(\tau') \\
\times \Bigl[ \chi_0^{-1}(\tau-\tau') - \mathcal{K}_c(\tau-\tau') \Bigr]\biggr\}
\end{multline}
and the trace is over the spin degrees of freedom.
$\chi_0^{-1}(\tau\!-\!\tau')$, with short-time cutoff $\tau_c=2\pi/\omega_c$,
encodes the bath spectrum. $\mathcal{K}_c (\tau\!-\!\tau')$, which has a Fourier
transform $\mathcal{K}_c (i\omega_n) = \kappa |\omega_n|$ with
$\kappa = \pi N_0 J_K$, comes from integrating out the fermions.

Trotter decomposition of $\mathcal{S}_{\text{imp}}'$ (with time slice $\tau_0$)
requires identifying the effect of the transverse-field term $\Gamma S^x$ with
that of a nearest-neighbor term in the corresponding Ising model described
\cite{Anderson.69+Yuval.70,Blume.70+Suzuki.76} by
\begin{equation}
\label{classicalIsing}
\mathcal{Z} \sim \text{Tr} \exp \Biggl[ \, \sum_{i=1}^L K_{\text{nn}}
  S_i^z S_{i+1}^z + \sum_{i,j=1}^L K_{\text{lr}}(i-j) S_i^z S_j^z \Biggr] .
\end{equation}
The nearest-neighbor interaction $K_{\text{nn}}=-2\ln(\tau_0\Gamma/2)$,
while the long-range interaction $K_{\text{lr}}(i-j)$ results from discretizing
$\chi_0^{-1}(\tau-\tau') - \mathcal{K}_c(\tau-\tau')$.
The bosonic spectrum specified by Eq.~(\ref{EQ:sub-Ohmic}) gives rise for
$\mbox{max}(\tau_0,\tau_c)\ll \tau<\beta/2$ to a simple power-law behavior
along the imaginary-time axis:
\begin{equation}
\chi_0^{-1}(\tau) \sim 1/|\tau|^{2-\epsilon} \qquad
   \text{for} \;\; 0 \le \epsilon < 1.
\end{equation}
The inverse temperature $\beta=1/T$ in the BFKM sets the length $L=\beta/\tau_0$
of the Ising chain, with the periodic boundary condition $S^z_{L+1}=S^z_1$
enforced by the trace operation.
Varying the bosonic coupling $g$ amounts to changing the effective temperature
$\mathcal{T}$ of the classical model (which bears no relation to $T$ in the
BFKM).

Classical Ising chains with this type of long-ranged interaction have been
studied for decades, and it is well established that the phase transition for
$\epsilon = 0$ is Kosterlitz-Thouless-like
\cite{Anderson.69+Yuval.70,Chakravarty.95} and is described over the range
$0 < \epsilon < 1$ by a local $\phi^4$ theory \cite{Fisher.72,Suzuki.72}. For
$0 < \epsilon < 1/2$ the phase transition is controlled by the interacting
Ginzburg-Wilson-Fisher fixed point, whereas for $1/2 < \epsilon < 1$
mean-field behavior obtains.

In general, if the classical system is below its upper critical dimension,
then close to the critical temperature ${\mathcal{T}}_c$ the only relevant scale is the finite
system size, and the static part of the order-parameter correlation function
should scale as $\chi_{\text{static}}(L)=L^x Y(L/\xi_{\infty})$, where
$\xi_{\infty} \equiv \xi_{\text{static}}(L=\infty) \sim
(\mathcal{T}-\mathcal{T}_c)^{-\nu}$. Combined with
$\chi_{\text{static}}(L=\infty) \sim (\mathcal{T}-\mathcal{T}_c)^{-\gamma}$,
we have $x=\gamma/\nu$. From the hyperscaling relation $\gamma=(2-\eta)\nu$
and the fact \cite{Fisher.72} that $\eta=1-\epsilon$, we end up with
$x=1-\epsilon$: at $\mathcal{T}=\mathcal{T}_c$,
\begin{equation}
\label{below}
\chi_{\text{static}}(L)\sim L^{1-\epsilon} \qquad
   \text{for} \;\; 0<\epsilon<1/2.
\end{equation}
Above the upper critical dimension, the finite-size scaling is complicated
by the presence of dangerously irrelevant variables that introduce additional
scales to the problem and destroy hyperscaling \cite{Brezin.82}.
Extensive theoretical work \cite{Brezin.82,Brezin.85+Luijten.96} has concluded
that at $\mathcal{T}=\mathcal{T}_c$,
\begin{equation}
\label{above}
\chi_{\text{static}}(L)\sim L^{1/2} \qquad \text{for} \;\; 1/2<\epsilon<1.
\end{equation}

Beyond the static limit, the full scaling properties of the susceptibility have
not been clarified. We are aware only of a finding (Fig.\ 4 of \cite{Luijten.97})
that $\chi(\tau,L)$ fails to show any scaling collapse in terms of $\tau/L$.
Separately, a study of the phase diagram has hinted at the importance of the
winding of the interaction around the finite-size Ising ring \cite{Cannas.04}.
These considerations motivate us to carry out a careful analysis of the scaling
of $\chi$ with particular attention to the effects of winding.

\textit{Finite-size scaling of the classical Ising
chain.}---The interaction $\chi_0^{-1}(\tau-\tau')$ is specified by
$\tilde{\chi}_0^{-1}(\omega) \equiv \text{Im} \chi_0^{-1}(\omega+i0^+)
= \pi g^2\omega^{1-\epsilon} \sgn(\omega)$, which follows from
Eq.~(\ref{EQ:sub-Ohmic}) with the cutoff $\omega_c$ taken to infinity:
\begin{align}
\label{long-range}
\chi_0^{-1}(\tau-\tau')
&= \int_0^{\infty}\!\! d\omega \, \tilde{\chi}_0^{-1}(\omega) \,
   \frac{\cosh\big[\omega(\beta/2-|\tau\!-\!\tau'|)\big]}{\sinh(\omega \beta/2)}
   \nonumber \\
&\equiv \chi_0^{-1}(\tau-\tau',T,\infty),
\end{align}
\mbox{where}
\begin{align}
\label{zetafkt}
\chi_0^{-1}(\tau,T,M)
& = \pi \Gamma(2-\epsilon) g^2 \sum_{n=0}^{M-1} \Bigl[
\bigl( |\tau| + n\beta \bigr)^{-(2-\epsilon)} \big. \nonumber\\
& \qquad + \big. \bigl( (n+1)\beta - |\tau| \bigr)^{-(2-\epsilon)}
  \Bigr].
\end{align}

We study the Ising chain with effective spin-spin interaction
$\chi_0^{-1}(\tau-\tau', T, M)$ for different values of
$M$, the maximum number of times that the interaction wraps around
the Ising ring of length $L$. The $|\tau-\tau'| \rightarrow \infty$ behavior is
not altered by choosing a finite value of $M$.
We use a cluster-updating Monte Carlo scheme \cite{Luijten.95,Wolff.98} to
measure the dynamical susceptibility $\chi(\omega_n,T)$
at coupling $g=g_c$ (i.e., $\mathcal{T}=\mathcal{T}_c$ in the classical model);
$\omega_n=2\pi n/(\tau_0 L)$, $n=0$, $\pm 1$, $\pm 2$, $\ldots$ are the
Matsubara frequencies of the quantum model. A binning analysis of our data shows
that the relative error $\Delta \chi/\chi \lesssim 10^{-2}$ and the integrated
autocorrelation time is small.

\begin{figure}[t!]
\includegraphics[width=0.96\columnwidth]{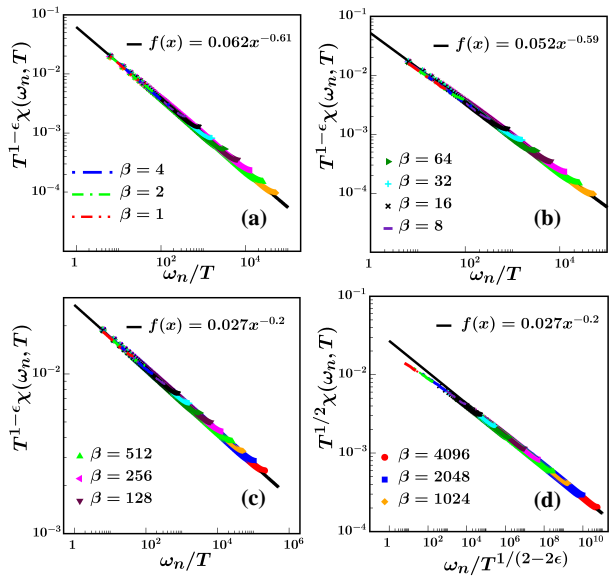}
\caption{\label{scalingeps0408}(Color online)
Finite-size scaling of the susceptibility below ($\epsilon=0.4$, upper
panels) and above ($\epsilon=0.8$, lower panels) the upper critical dimension,
for $\tau_0=1/16$, $\kappa=K_{\text{nn}}=0$, and winding numbers
(a,c) $M=1$ and (b,d) $M=10^9$. For $\epsilon>1/2$,
the temperature scaling depends on the choice of $M$. Similar
results have been obtained for various $\tau_0$ between $1/4$ and $1/64$.}
\end{figure}

For $0<\epsilon<1/2$, the scaling properties do not depend on the choice
of $M$ and the static susceptibility obeys Eq.\ (\ref{below}).
As a result, the dynamic susceptibility scales as
\begin{equation}
\label{chi:omega-over-T}
\chi(\omega_n,T)=T^{-(1-\epsilon)}\Phi(\omega_n/T).
\end{equation}
Figure \ref{scalingeps0408} illustrates this $\omega_n/T$ scaling for
$\epsilon=0.4$ with $M=1$ [part (a)] and $M=10^9$ [part (b)].

For $1/2<\epsilon<1$, by contrast, the scaling properties depend crucially
on $M$, as exemplified in Fig.~\ref{scalingeps0408} for
$\epsilon=0.8$. The scaling plot for $M=1$ [part (c)] implies
that $\chi(\omega_n) \sim |\omega_n|^{-0.2}$ and $\chi(T) \sim T^{-0.2}$.
Only for $M\to\infty$ do we recover the anticipated mean-field
result of Eq.\ (\ref{above}), as shown in Fig.\ \ref{scalingeps0408}(d) for
$M=10^9$. Between these extremes, there is a slow crossover with
increasing $M$, exemplified by the evolution of
$\chi_{\text{static}}(T)$ \cite{version1}. Difficulty in recovering mean-field
physics in this model has been observed in the context of the phase diagram
\cite{Cannas.04}, but the origin of this behavior and its implication for
$\omega_n/T$ scaling have not previously been recognized.

\begin{figure}[t!]
\includegraphics[width=0.95\columnwidth]{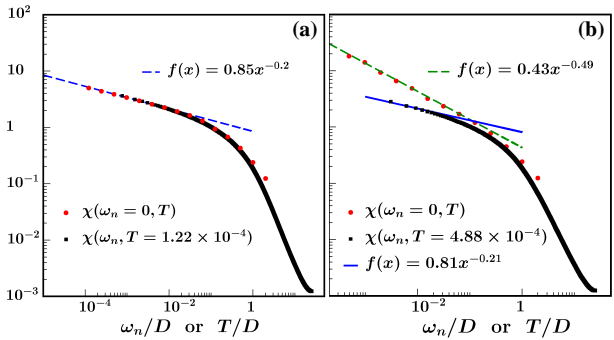}
\caption{\label{finite_kc}(Color online)
Static and dynamical susceptibilities for $\epsilon=0.8$
with $\kappa=2$ and $K_{\text{nn}}$
specified by $\Gamma = 0.75$ and $\tau_0=1/8$:
(a) $M=1$, and (b) $M=5 \times 10^7$.}
\end{figure}

In obtaining Fig.\ \ref{scalingeps0408}, we set $\kappa=0$ and
$K_{\text{nn}}=0$ because the construction of clusters via continuous bond
probabilities is efficient only for pure power-law interactions
\cite{Luijten.95}. For nonzero $\kappa$ and $K_{\text{nn}}$, the calculations
can be carried out only for a more limited range of $\omega_n$ and $T$, but the
same behavior is found, as illustrated in Fig.~\ref{finite_kc}.

The difference between the cases of infinite and zero winding can be
understood as follows. The infinite-winding interaction
$\chi_0^{-1}(\omega_n,T,M=\infty)$ depends only on $\omega_n$.
The temperature dependence of $\chi(\omega_n,T)$ is then entirely determined
by that of the ``spin self-energy''
\begin{eqnarray}
\label{self-energy}
\mathcal{M}(\omega_n,T) \equiv \chi_0^{-1}(\omega_n,T) - 1/\chi(\omega_n,T) .
\end{eqnarray}
The temperature dependence of $\mathcal{M}$ is controlled by the dangerously
irrelevant quartic coupling of the local $\phi^4$ theory, and is proportional
to $T^{1/2}$. On the other hand, the zero-winding interaction
$\chi_0^{-1}(\omega_n\!=\!0,T,M\!=\!1) \sim T^{1-\epsilon}$,
so $\chi^{-1}(T,\omega_n)$ acquires temperature dependence from both
$\chi_0^{-1}$ and $\mathcal{M}$. The leading term varies as $T^{1-\epsilon}$,
which overpowers the $T^{1/2}$ from the dangerously irrelevant coupling,
so the $\omega/T$ form of Eq.\ (\ref{chi:omega-over-T}) ensues.
We have confirmed that this distinction between $M=1$ and
$M=\infty$ is robust over the range $1/64\le \tau_0 \le 1/4$.

\begin{figure}[t!]
\includegraphics[width=.7\columnwidth]{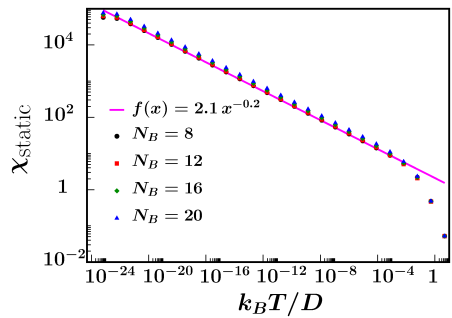}
\caption{\label{BFKM}(Color online)
Static susceptibility at the QCP of the Ising BFKM for $\epsilon=0.8$,
calculated using the NRG with discretization $\Lambda=9$ and various values
of the bosonic truncation parameter $8\le N_b\le 20$.
See \protect\cite{Glossop.05} for further details.}
\end{figure}

\textit{Quantum critical behavior of the BFKM.}---We have systematically
extended previous NRG calculations \cite{Glossop.05} to investigate the
robustness of the quantum critical behavior. Figure \ref{BFKM} shows data
for $\epsilon=0.8$ that can be fitted to $\chi_{\text{static}}(T) \sim T^{-x}$
over more than 20 decades of $T$ with $x=1-\epsilon$ to better than
1\% accuracy. The value of the exponent shows no systematic evolution as the
bosonic truncation parameter $N_B$ \cite{Glossop.05} increases from 8 to 20.

These NRG results, which are consistent with $\omega/T$ scaling, are in stark
contrast with the $\omega/T^{1/(2-2\epsilon)}$ scaling and exponent $x=1/2$
found in the regime $1/2<\epsilon<1$ of the classical Ising chain (with
infinite winding). This disparity signals the failure of the
quantum-to-classical mapping for the QCP of the Ising BFKM.

The most likely source of the breakdown lies in the replacement of the spin-flip
(transverse Kondo-coupling) term in the quantum model by a
$(\partial_\tau \phi)^2$ term in the local $\phi^4$ theory. In the BFKM, spin
flips are essential for the formation of the Kondo-screened state and, by
extension, central to the nature of the QCP displaying a critical Kondo effect.
(Spin flips are the one-dimensional analog of the vortices in, for instance,
the short-range XY model in two dimensions.) In the local $\phi^4$ theory,
however, the $(\partial_\tau\phi)^2$ term (\textit{i.e.}, the
$\omega_n^2\phi^2$ coupling) is overtaken at low frequencies by the
$|\omega_n|^{1-\epsilon} \phi^2$ term associated with the long-range
interactions of non-spin-flip type.

To expand on this point, the short-range interaction term in the discrete Ising
chain [Eq.\ (\ref{classicalIsing})] that arises from mapping the transverse
Kondo coupling can be expressed
\begin{equation}
\label{vortex}
\tau_0^2 \, \log \left( \frac{\Gamma \tau_0}{2} \right)
  \sum_{i=1}^L \frac{(S^z_{i+1}-S^z_i)^2-\frac{1}{2}}{\tau_0^2} .
\end{equation}
In the continuum limit, the corresponding kinetic energy written in terms of
$(\partial S^z/\partial \tau)^2$ clearly requires regularization through a
finite value of $\tau_0$. The long-range interaction in Eq.\
(\ref{classicalIsing}) must also be regularized, either by reinstating
$\omega_c$ in $\tilde{\chi}_0^{-1}(\omega)$ entering Eq.\ (\ref{long-range}) or
through enforcement of $|\tau-\tau'|>\tau_c$ in Eq.\ (\ref{zetafkt}). Finally,
$\tau_0$ serves to regularize the dynamically generated Kondo scale,
$T_K \approx (1/\tau_0)\exp(-1/\Gamma\tau_0)$, below which the quantum scaling
occurs. This scale vanishes when $\tau_0$ (or $\tau_c$) vanishes, making the
quantum critical behavior inaccessible. These observations suggest
that the continuum limit, necessarily taken in the quantum-to-classical
mapping, fails to capture the the topological (vortex-like) effect
encoded in the Kondo spin-flips. What results is a change in the temperature
dependence of the spin self-energy, qualitatively similar to that arising in
the classical model from changing the maximum winding from $M=1$ to $M=\infty$.

A concurrent study \cite{Winter.09} addresses the mapping of the dissipative
spin-boson model using a Monte Carlo algorithm that explicitly takes the limit
$\tau_0 \rightarrow 0$ of a classical Ising chain \cite{Rieger.93}. 
Ref.\ \cite{Winter.09} focuses on the finite-size
scaling exponents of the static magnetic properties, rather than the scaling
of the dynamical susceptibility.
Its results for the mapped classical action are compatible with, and
complementary to, our own. Ref.~\cite{Winter.09} also speculates that a
marginal coupling introduced by bosonic truncation invalidates the NRG results
\cite{Vojta.05} for the spin-boson model. However, our NRG studies for the
BFKM, carried out over an extended range of bosonic truncation parameters
and covering more than 20 decades of temperature, show evidence, neither
for evolution in the exponent $x$ away from its interacting value of
$1-\epsilon$ towards the mean-field $x=\frac{1}{2}$, nor for the line of
critical points that would be expected in the presence of a marginal operator.
We also note that the NRG conclusion for the BFKM that $x=1-\epsilon=y$ cannot
be vitiated by a dangerously irrelevant coupling. The NRG result corresponds
to a spin self-energy [see Eq.\ \eqref{self-energy}]
$\mathcal{M}(T) = \chi_0^{-1} - 1/ \chi \sim T^{1-\epsilon}$. In the regime
$\frac{1}{2}<\epsilon<1$ of interest, this temperature dependence dominates
any $T^{1/2}$ term potentially generated by a dangerously irrelevant coupling,
in a manner reminiscent of what happens in the Monte Carlo calculations with
no winding ($M=1$). These considerations all point to the quantum-to-classical
breakdown being a real phenomenon.

In summary, we have addressed the quantum-to-classical mapping of the Ising
Bose-Fermi Kondo impurity problem. The finite-size scaling for the spin
susceptibility of the mapped classical chain demonstrates an intriguing
dependence on the winding of the long-ranged interaction under periodic boundary
conditions. Only for infinite winding does one recover the expected mean-field
behavior. The contrast between these scaling properties and those of the
Bose-Fermi Kondo model suggests a breakdown of the quantum-to-classical
mapping for Ising-anisotropic quantum dissipative systems
arising from the manner in which Kondo spin flips are treated in the
continuum limit that is taken in such a mapping.

We thank C.\ J.\ Bolech, M.\ Glossop, H.\ Rieger, T.\ Vojta, and S.\ Yamamoto
for useful discussions, and M.\ Troyer for generously sharing with us his
unpublished work with M.\ Guidon and P.\ Werner. This work has been supported
in part by the NSF  Grant No.
DMR-0706625 (S.K. and Q.S.) and the NSF Grant No. 0710540 (K.I.), 
the Robert A.\ Welch Foundation, the W.\ M.\ Keck
Foundation, the Rice Computational Research Cluster funded by the NSF and
a partnership between Rice University, AMD, and Cray.


\end{document}